# Modeling of the Labour Force Redistribution in Investment Projects with Account of their Delay

I.D. Kolesin[2], O.A. Malafeyev[2], I.V. Zaitseva[*1], A.N. Ermakova[1], D.V. Shlaev[1]

[1]Stavropol State Agrarian University, 12, Zootechnicheskiy Lane, Stavropol, Russia, 355017

[2]Saint-Petersburg State University, 7-9, University Emb. , Saint-Petersburg, Russia, 199034

[*1]zirinazirina2015@yandex.ru, [1]dannar@list.ru, [1]shl-dmitrij@yandex.ru, [2]malafeyevoa@mail.ru

September 19, 2017

**Abstract.** The mathematical model of the labour force redistribution in investment projects is presented in the article. The redistribution mode of funds, labour force in particular, according to the equal risk approach applied to the loss of some assets due to delay in all the investment projects is provided in the model. The sample of the developed model for three investment projects with the specified labour force volumes and their defined unit costs at the particular moment is given.

## 1. Introduction

The introduction of investment projects in volatile conditions is followed by the risk to lose the major part of the assets due to the investments delay. These losses are stipulated by, for instance, inflation. In this case, having received the initial sum of money, a company could pay for some equal (without priority) work or resources. As the inflation level varies for different resources, it is reasonable to follow the principle of equity and allot some assets for every investment projects with the equal loss risk due to delay. Since all the investment projects are equal, the terms of delay is also supposed to be equal [1-3].

It should be noted that the principle of equity allows the company avoiding the conflict with a customer if it hasn't apply any other approach.

We'll define the problem of the specified assets redistribution per labour force following the risk equity [4-7].

## 2. The Mathematic Model of the Labour Force Redistribution in Investment Projects

### 2.1. Problem Setting

$n$ of investment projects are considered. It is required to allot the resources (as well as labour force) in volume $V_1, \ldots V_n$ in $n$ investment projects. Suppose, this work requires $S_i = c_i V_i, i = \overline{1,n}$, $S_1, \ldots, S_n$). But the assets at present *(B)* are not enough for the labour force payment in $n$ all the investment projects, i.e.
$$S_1 + S_2 + \cdots + S_n > B.$$

If only a part of labour force is employed in an investment project, it could require more investing till the further funds are received because of inflation. Suppose, at present the labour resources are divided in volumes
$$(c_1 u_1 + \cdots + c_n u_n = B).$$
$u_1, \ldots, u_n$

Then, the following investments are required to complete the balance:
$$S'_i = c'_i(V_i - u_i),\ c'_i > c_i,\ i = \overline{1,n},$$
where, $c_i$ – costs per unit at the time $t_0 = 0$, $c'_i$ - costs per unit after the time $t$. Let us assume costs growth pro rata time $t$:
$$c'_i = c_i + k_i t,\ k_i > 1,\ i = \overline{1,n}.$$

We comprise the ratio:
$$S'_i/S_i = r_i,\ \overline{1,n}.$$

We call $r_i$ $i$- investment project failure risk.

It is required to allot the assets (B) according to the equal risk approach in all the investment projects.
$$r_1 = \cdots = r_n,$$
i.e. to find such $u_1, \ldots, u_n$, that the formula
$$\frac{c'_i(V_i - u_i)}{c_i u_i} = r$$
is fulfilled and, where (r is unknown value) [8-10].

### 2.2. Method of Solution

Suppose, the terms of delay $T_1, \ldots, T_n$ are set. Then, $c'_i = c_i + k_i T_i,\ i = \overline{1,n}$.

Comprising the ratio $\frac{c_i(V_i - u_i)}{c_i u_i} = r_i, i \in \overline{1,n}$, we find the formula for $u_i$ out of every expression
$$u_i = \frac{c'_i V_i}{c'_i + c_i r_i}, i \in \overline{1,n}.$$

Let us substitute these formulas to the equation $c_1 u_1 + \cdots + c_n u_n = B$ and suppose $r_i = r, i \in \overline{1,n}$. We get the equation as to the unknown $r$. Suppose, its solution is $r = r$. Substituting the value found $r$ to the expression for $u_i$, we find all $u_i$ [11-14].

### 2.3. Sample

Suppose, there are three investment projects (n) with labour force volumes: $V_1 = 100, V_2 = 300, V_3 = 250$ and costs per unit at the time $t_0 = 0$ are $c_1 = 2, c_2 = 3, c_3 = 1$.

Suppose, all the works are delayed at the time $T_1 = T_2 = T_3 = 10$, the labour force inflation rises respectively to time, and the proportionality coefficients are defined:
$$k_1 = 0.1, k_2 = 0.4, k_3 = 0.2.$$

Then in time T=10, the new costs per unit are:
$$c'_1 = 3, c'_2 = 7, c'_3 = 3\ (c'_i = c_i + k_i T_i, i = 1,2,3).$$

Suppose, the company obtains the assets in the amount of *B=295* at the time $t_0 = 0$, whereas, the projecting costs of assets is
$$2 \cdot 100 + 3 \cdot 300 + 1 \cdot 250 = 1350.$$

It is necessary to allot the assets B in three investment projects, so that the total cost of labour force in volumes $u_1, u_2, u_3$ is equal to *B=295*:
$$c_1 \cdot u_1 + c_2 \cdot u_2 + c_3 \cdot u_3 = 295.$$

Then we comprise the expression based on the equal risk approach:
$$\frac{c'_1(100 - u_1)}{c_1 u_1} = \frac{c'_2(300 - u_2)}{c_2 u_2} = \frac{c'_3(250 - u_3)}{c_3 u_3} = r.$$

Having expressed each of $u_1, u_2, u_3$ in r and substituted them in *(*)*, we receive:
$$c_1 \frac{c'_1 100}{c'_1 + c_1 r} + c_2 \frac{c'_2 300}{c'_2 + c_2 r} + c_3 \frac{c'_3 250}{c'_3 + c_3 r} = 295.$$

The only positive root *r* of this equation will be equal to *r\*=8.3* (residual 295-294.476=0.524).

Substituting *r\** in the expression for $u_i$
$$u_i = \frac{c'_i V_i}{c'_i + c_i r}, \; i = 1,2,3,$$
we find the required volumes of labour force:
$$u_1 = 15.3, u_2 = 65.8, u_3 = 66.4.$$

## 3. Results and Discussion

It should be mentions that theoretical and practical aspects of the sustainable growths of the economy, in general, and the economy of particular regions are not worked out sufficiently. At present, there is much concern about the labour force management. If the essence of the labour force management is clear enough, its basic principles are to be identified and studied.

Thus, the topicality of labour force redistribution for providing the stability and balance of positive geographical shifts is stipulated by the necessity to study social and economic factors of labour force development in the current economic environment.

The authors have studied the essence and performance of labour force in reasonable detail [15-16]. The problems concerning the labour force management on the basis of its structure were sufficiently considered, the main philosophy of formation management and major mathematical model-based methods were identified [17-18]. The economic methods of analysis, assessment and forecasting of the regional labour potential used to study the labour sector in agriculture Stavropol Territory were introduced [19-20].

## 4. Conclusion

In view of the above-mentioned, the research of the principles applied for the labour force alteration, the assessment of labour force planning and its impact on the regional economic development, the development and justification of methods and procedures used in the labour force management are turned out to be topical.

The concept of economy development based on the analysis considering the assessment and forecasting the labour force level could be formed as the result of this research. The solution of the complex problems connected with the labour force management refers to the formulation of the new theoretical and methodological approach to the management system. Therefore, it's required to create the corresponding economic and mathematic modeling apparatus, management and optimization procedures, to determine the quality criteria for transition processes and perspective management laws. The reference model differs by the ability to achieve the qualified take-off and immediate intellectual development of labour force at work getting more and more complex should be taken as a base.